\documentclass[aps,prd,nofootinbib,reprint,superscriptaddress,longbibliography]{revtex4-2}
\usepackage{caption}
\usepackage{etoolbox}
\usepackage{dcolumn}% Align table columns on decimal point
\usepackage{bm}% bold math
\usepackage{graphics}
\usepackage{subcaption}
\usepackage{xcolor}
\usepackage{dirtytalk}
\usepackage{appendix}
\usepackage{graphicx} %Include figure filesusepackage{graphicx} %Include figure files
\usepackage{blindtext}
\usepackage{amsmath}
\usepackage{amsfonts}
\usepackage{amssymb}
\usepackage{mathtools,halloweenmath}
\usepackage[%
colorlinks=true,
urlcolor=blue,
linkcolor=red,
citecolor=blue
]{hyperref}
\usepackage{physics}
\NewDocumentCommand{\tens}{t_}
{%
	\IfBooleanTF{#1}
	{\tensop}
	{\otimes}%
}
\NewDocumentCommand{\tensop}{m}
{%
	\mathbin{\mathop{\otimes}\displaylimits_{#1}}%
}
\usepackage{natbib}
\begin{document}
	\title{\bf New insights on mutual information in the island approach to the Page curve}
	\vskip 1cm
    \author{Anirban Roy Chowdhury}
	\email{iamanirban@bose.res.in}
	\affiliation{Department of Astrophysics and High Energy Physics,\linebreak
		S.N.~Bose National Centre for Basic Sciences,\linebreak
		JD Block, Sector-III, Salt Lake, Kolkata 700106, India}
	\author{Souvik Paul}
	\email{souvik.paul@bose.res.in}
	\affiliation{Department of Astrophysics and High Energy Physics,\linebreak
		S.N.~Bose National Centre for Basic Sciences,\linebreak
		JD Block, Sector-III, Salt Lake, Kolkata 700106, India}	
	\author{Sunandan Gangopadhyay}
	\email{sunandan.gangopadhyay@bose.res.in}
	\affiliation{Department of Astrophysics and High Energy Physics,\linebreak
		S.N.~Bose National Centre for Basic Sciences,\linebreak
		JD Block, Sector-III, Salt Lake, Kolkata 700106, India}	
	
	\begin{abstract}
		\noindent In this article, we have presented one of the very important observations, regarding the behavior of mutual information of two different sets of subsystems in the after Page time scenario. This provides us with some deep insights about the redistribution of geometrical correlation between the two different sets of subsystems on a Cauchy slice. In our earlier works, we have shown how the saturation of mutual information between two specific subsystems plays a vital role in obtaining the correct Page curve for the eternal black hole. In those works, we have shown that the mutual information between $B_+$ and $B_-$, that is, $I(B_{+}: B_{-})$, vanishes at scrambling time, which leads to the correct Page curve. That means that at scrambling time, there is no correlation between $B_+$ and $B_-$. Remarkably, it is observed that at this particular value of observer's time, the mutual information between $\mathcal{I}$ and $R$, that is, $I(\mathcal{I}:R)$, becomes singular. This indicates that the regions $\mathcal{I}$ and $R$ become maximally entangled. This provides us with a notion of the transfer of geometric correlation between different regions on the Cauchy slice. In this work, we have also provided a way to calculate the tripartite mutual information of regions $\mathcal{I}$,~$R_+$ and $R_-$, that is, $I(\mathcal{I}:R_+:R_-)$ on the Cauchy slice using the earlier results involving the bipartite regions. This is a new result which was missing in the earlier literature.
	\end{abstract}
	\maketitle
\section{Introduction}
\noindent In recent years, the black hole information loss paradox has become one of the most fascinating problems to theoretical physicists. This provides a bridge between quantum mechanics and general relativity. To understand this problem, we need a complete quantum theory of gravity. But we still do not have a consistent theory of quantum gravity. This black hole information loss problem was first introduced by Hawking. Hawking shows that an entangled pair of particles is created due to the vacuum fluctuation of the quantum field near the black hole horizon \cite{Hawking:1975vcx}. One particle of this pair falls into the black hole, and the other one escapes to infinity. This outgoing particle is known as the Hawking quanta. The whole analysis made by Hawking was semi-classical, but years of research in this direction show that this radiation is quantum mechanical. Therefore, it is interesting to investigate various quantum mechanical observables. In particular, the study of the von-Neumann entropy of Hawking radiation gives rise to a serious problem. The behavior of the von-Neumann entropy of Hawking radiation suggests violation of unitarity in quantum mechanics. This is known as the information loss paradox. One can explain this paradox in the following way. It was known that a black hole is formed due to the collapse of a massive shell that is in a pure state. Therefore, the von-Neumann entropy of radiation at the beginning of the black hole formation is zero. Therefore, the unitary principle suggests that at the end of the evaporation, the final state of the radiation must be in the pure state. This means that at the end of the evaporation, the von-Neumann entropy of the black hole radiation should vanish. However, Hawking's analysis shows that, the von-Neumann entropy black hole radiation monotonically increases with time, and it does not vanish after the black hole evaporation \cite{PhysRevD.14.2460,PhysRevD.13.191}. This observation suggests the violation of unitarity.\\
There is also another way to understand this paradoxical situation from Page curve \cite{PhysRevLett.71.3743}. The Page curve describes the time evolution of the von Neumann entropy of Hawking radiation. Therefore, one can argue that, this paradox can be solved if one can obtain the correct Page time of Hawking radiation. Before explaining the Page curve, we should keep in mind that the von-Neumann entropy of the Hawking radiation is identified with that of the matter fields residing on the outside region of the black hole, namely the region $R=R_+\cup R_-$ \footnote{It is also to be noted that the von-Neumann entropy is also known as the fine-grained entropy of the system.}. According to Page, starting from zero, the von-Neumann entropy of the Hawking radiation ($S_{vN}(R)$) increases with the observer's time. On the other hand due to the evaporation, the thermodynamic entropy of the black hole ($S_{BH}$\footnote{The thermal entropy is also known as the coarse-grained entropy.}) \cite{Bekenstein:1973ur} decreases with time. Then there is a particular value of the observer's time, which is known as the Page time ($t_{P}$), at which the von-Neumann entropy of the black hole gets equal to the thermal entropy of the black hole. It is also to be noted that, up to the Page time ($t\le t_{P}$), there is no paradoxical situation because in this time domain the von-Neumann entropy of the black hole is smaller than the thermal entropy of the black hole. However after the Page time, that is, when $t>t_{P}$, the von-Neumann entropy of Hawking radiation is greater than the thermal entropy of the black hole. Therefore, the paradoxical situation arises just after the Page time. Therefore, the challenge is to obtain the correct form of Page curve after the Page time. Therefore, according to Page to maintain the unitarity during the complete process (from black hole formation to complete evaporation), the time evolution of the von-Neumann entropy should have the following nature. At the beginning of the evaporation, the $S_{vN}(R)$ is zero, then as time passes, the von-Neumann entropy of the radiation increases with time and then at Page time $(t=t_{P})$, the von-Neumann entropy of the radiation becomes equal to the thermal entropy of the black hole. Now to maintain the unitarity $S_{vN}(R)$ should decrease with time and finally at the end of the evaporation the $S_{vN}(R)$, should vanishes. On the other hand for an eternal black hole the von-Neumann entropy of the radiation should be constant with time. There are some very interesting observation regarding this which can be found in \cite{Page_2013,Almheiri:2012rt}.\\
The above discussion suggests that one need to obtain a correct form of the Page curve after the Page time. Recent developments in AdS/CFT correspondence \cite{Maldacena:1997re,Witten:1998qj,Aharony:1999ti,Almheiri:2019yqk} and the idea of entanglement wedge reconstruction provide us with a very nice way to compute the fine-grained entropy of the Hawking radiation. This formula of computing the von-Neumann entropy of Hawking radiation is known as the ``\textit{island formula}" \cite{Penington:2019kki,Penington:2019npb,Almheiri:2019hni}. According to this formula there is a certain region inside the black hole, which is named the island, that contributes to the fine-grained entropy of the Hawking radiation. It is to be noted that the endpoints of the island is known as the quantum extremal surfaces (QES)\cite{Engelhardt:2014gca,Hubeny:2007xt,PhysRevLett.96.181602,PhysRevLett.96.181602}. Keeping this additional contribution in mind, one can write down the formula of computing the fine-grained entropy of Hawking radiation in the following manner
\begin{equation}\label{eq1}
S(R)=\min \underset{\mathcal{I}}{\text{ext}} \left\{\frac{A(\partial \mathcal{I})}{4G_N}+S_{vN}(\mathcal{I}\cup R)\right\}~.
\end{equation}
One can derive the above formula by applying the replica trick in a gravitational background \cite{Almheiri:2019qdq,Penington:2019kki,Goto:2020wnk,Colin-Ellerin:2020mva}. This method suggests that to compute the fine-grained entropy of the radiation, we need to consider $n$ copies of the black hole system and glue them appropriately (with suitable boundary conditions). Then we need to compute the partition function in that replicated geometry and compute the Rényi entropy. Then, in a certain limit, we get the fine-grained entropy of the Hawking radiation. In this method of computing the fine-grained entropy, we have encountered two different saddle points in the partition function. One is formally known as the Hawking saddle point, which dominates in the past time domain. When the Hawking saddle dominates, there is no island contribution, which results in a monotonically increasing nature of fine-grained entropy of black hole radiation. This is nothing but the results given by Hawking. On the other hand, in the post Page time era, we have another non-trivial saddle point, which is known as the replica wormhole saddle point. In this time domain, the island starts to contribute to the fine-grained entropy of black hole radiation.  This formula gives us a correct time evolution of the fine-grained entropy of Hawking radiation, which obey the unitarity. There are some interesting works in this direction that can be found in \cite{Almheiri:2020cfm,chen2020information,Dong:2020uxp,Anegawa:2020ezn,Balasubramanian:2020xqf,Raju:2020smc,Alishahiha:2020qza,Azarnia:2021uch,Arefeva:2021kfx,He:2021mst,Omidi:2021opl,Yu:2021rfg,Yadav:2022fmo,Du:2022vvg}.\\
The \textit{island formula} was first applied to a black hole in JT gravity filled with two-dimensional conformal filed theory \cite{Goto:2020wnk,Hollowood:2020cou}. It was shown that in the pre-page time scenario, that is, when $t<t_P$, the Hawking saddle point dominates in the gravitational path integral. This leads to a monotonically increasing behaviour of the fine-grained entropy of Hawking radiation. Subsequently, at $t=t_P$, the fine-grained entropy of the radiation gets equal to the coarse-grained entropy of the black hole. And just after the Page time starts to dominate and the contribution from the
interior region $(\mathcal{I} )$ becomes significant, which eventually lead
us to the expected Page curve. Owing to the remarkable features of the ``\textit{island formula}," it has garnered significant attention from researchers. Some recent works in this direction can be found in \cite{Hartman:2020swn,Wang:2021mqq,Li:2021lfo,Pedraza:2021cvx,Almheiri:2019psy,Matsuo:2020ypv,Krishnan:2020fer,Krishnan:2020oun,Goswami:2022ylc,Ling:2020laa,Kim:2021gzd,Ahn:2021chg,Bhattacharya:2021jrn,Chen:2020tes,Jain:2023xta,Guo:2023fly,Chang:2023gkt,Li:2023fly,Guo:2023gfa,Miao:2023unv,Yadav:2022jib,HosseiniMansoori:2022hok,Hu:2022zgy,Yu:2022xlh,Djordjevic:2022qdk,Anand:2022mla,Azarnia:2022kmp,Seo:2022ezk,Yu:2023whl,Uhlemann:2021nhu,Yu:2025euq,Cheng:2025bnj,Saito:2025lsv,Wang:2024itz,Yu:2024fks,Goswami:2023ovb,Zhong:2024fmn,Hartman:2020khs,Espindola:2022fqb,Ben-Dayan:2024iwc,Ben-Dayan:2022nmb,Basu:2025sqk,Geng:2021wcq,Hao:2024nhd,Geng:2024xpj,Anous:2022wqh,Ghosh:2021axl,Afrasiar:2022ebi,Xiao:2026cla,Li:2026lxa}.\\
In our earlier works \cite{Saha:2021ohr,RoyChowdhury:2022awr,RoyChowdhury:2023eol,Saha:2025ttj,RoyChowdhury:2026jjp}, we have provided an alternative idea to obtain the correct form of Page curve instead of following the standard technique which are provided earlier. We have addressed this problem from the idea of mutual information. Our work suggests that the mutual information plays a vital role in determining the correct form of the Page curve in the post-Page time domain for eternal black holes. In particular, we have shown that in the presence of an island, the mutual information between $B_+$ and $B_-$, that is $I(B_+:B_-)$, vanishes at the ``\textit{scrambling time}'' $t_{Scr}$. This surprisingly gives a time independent expression of matter field entropy (given by the second term in eq.\eqref{eq1}). Then, adding the area term with the matter field entropy and following the standard process of extremization, we get the position of the island endpoints and the time-independent expression of the fine-grained entropy of Hawking radiation. This yields the correct form of Page curve in the post Page time domain. We have shown that our formalism holds good for eternal black holes in asymptotically flat, de Sitter and Anti-de Sitter spacetime.\\
In this work, we have reported two important findings. Our former work suggests that just after the inclusion of the island, the mutual information between $B_+$ and $B_-$ vanishes when $t_a-t_b=|r_{*}(a)-r_{*}(b)|$. This means there is no mutual correlation between $B_+$ and $B_-$. They are in the disconnected phase. On the other hand, we have shown that imposing the analogous condition of vanishing mutual information between $B_+$ and $B_-$ leads to a divergence in the mutual information between $\mathcal{I}$ and $R~(R\equiv R_+ \cup R_-)$; that is, $I(\mathcal{I}:R)$ diverges. It should be noted that all the regions, that is, $R_+$, $R_-$, $\mathcal{I}$, $B_+$ and $B_-$ are in the same Cauchy slice. Our observation suggests that under similar conditions $I(B_+: B_-)$ vanishes, but $I(\mathcal{I}: R)$ diverges. This suggests that there is a transfer of mutual correlation among the different sets of regions in the same Cauchy slice. This in turn implies that, under the analogous condition, the regions $B_+$ and $B_-$ are uncorrelated, whereas the regions $\mathcal{I}$ and $R$ become strongly correlated. It is 
%\begin{figure}
    %\centering
    %\includegraphics[width=0.7\linewidth]{withIsland.pdf}
   % \caption{Enter Caption}
    %\label{fig:placeholder}
%\end{figure}
an important observation. On the other hand this divergence of mutual information also leads us to get the correct expression of von-Neumann entropy of Hawking radiation in the post Page time domain. Thus, we get the correct form of Page curve for the eternal black hole. 
Thus, obtaining the expected Page curve necessitates one of two possibilities. Either the subsystems $B_{+}$ and $B_{-}$ are completely uncorrelated, $I(B_{+}:B_{-})=0$
or the island $\mathcal{I}$ is maximally correlated with the radiation region $R=R_{+}\cup R_{-}$, $I(\mathcal{I}:R)=\infty$. Hence, the correct Page curve can be realized only through either the complete absence of correlations between $B_{+}$ and $B_{-}$ or by requiring the maximal correlation between the $\mathcal{I}$ and the region $R=R_{+}\cup R_{-}$.
Therefore, to get the correct form of the Page curve for an eternal black hole, the two different mutual informations should obey the following condition simultaneously
\begin{equation}
I(B_+: B_-)=0~,~ I(\mathcal{I}:R)\rightarrow\infty~.
\end{equation}
%Thus we can argue that the c\\
This is one of the main observations in this work. The physical interpretation of this observation is as follows. Our analysis shows that the vanishing of the mutual information between the two exterior regions ($B_+$ and $B_-$), that is, $I(B_{+}:B_{-})$, and the singular behaviour of the island-radiation mutual information $(I(\mathcal{I}:R))$ occur under the same condition $(t_{a}-t_{b}=|r_{*}(a)-r_{*}(b)|)$. This leads us to conjecture that a nontrivial redistribution of the mutual information between regions on the Cauchy slice have taken place. We substantiate this conjecture by providing the plots of $I(B_+:B_-)$ and $I(\mathcal{I}:R)$ in Fig.\eqref{fig:comparison MI}.\\
Besides entanglement entropy and mutual information, another valuable measure of entanglement for a system composed of three separate parts is the tripartite information \cite{Kitaev:2005dm,Casini:2008wt}. In the classical information-theory context, it is also called the I-measure \cite{79902} or the interaction information \cite{mcgill1954multivariate}. For three disjoint subsystems $A,~B$ and $C$ the tripartite information is defined as
\begin{align}
    I(A:B:C)&=S(A)+S(B)+S(C)-S(A\cup B)\nonumber\\&-S(A\cup C)-S(B\cup C)+S(A\cup B \cup C)\nonumber\\&=I(A:B)+I(A:C)-I(A:B\cup C)~.
\end{align}
The tripartite information is a measure of the extensivity of the mutual information. The mutual information is extensive when $I(A:B:C)=0$, superextensive when $I(A:B:C)<0$ and subextensive when $I(A:B:C)>0$ \cite{Hayden:2011ag,Asadi:2018ijf}. For the extensive or the superextensive case, mutual information is said to be monogamous in nature. In general, quantum systems can have positive, negative or zero tripartite information. Although in \cite{Hayden:2011ag}, it was discussed that for holographic theories, mutual information is always monogamous, that is $I(A:B:C)\leq 0$. Some recent studies on the multipartite entanglement for various bulk spacetime backgrounds can be found in \cite{Asadi:2018ijf,Mirabi:2016elb,Ju:2023tvo,Ali-Akbari:2019zkf,Alishahiha:2014jxa,Ben-Ami:2014gsa,Ju:2024kuc,Maulik:2022hty,Asadi:2018lzr}. In the context of islands, the direct computation of the tripartite information between the island and radiation regions is quite challenging.
In this paper, we have provided %Another important finding of our work is thatwe have provided 
an alternative way to compute the tripartite mutual information among $\mathcal{I}$,~$R_+$ and $R_-$, that is, $I(\mathcal{I}: R_+: R_-)$. It is not possible to compute this tripartite mutual information directly from any technique. We have found that this tripartite mutual information is always negative with respect to both the island time parameter and the observer's time, that is, 
\begin{equation}
I(\mathcal{I}: R_+: R_-)<0~.
\end{equation}
This implies the fact that the tripartite mutual information among three disjoint regions on the Cauchy slice, namely, $\mathcal{I}$, $R_+$ and $R_-$ is always monogamous. On the other hand, this tripartite mutual information is also superextensive in nature. This provides us with some new insights regarding the distribution of geometric correlation among the three disjoint regions lying on the same Cauchy slice.
%mutuafine-grained entropy of the matter field, which is represented by the second term in eq.\eqref{eq1}. It is to be noted that, earlier, the matter field entropy, that is $S_{vN}(\mathcal{I}\cup R)$, is computed from the idea of pure state in the Cauchy slice. Earlier it was shown that $S_{vN}(\mathcal{I}\cup R)$ is computed by computing $S_{vN}(B_+\cup B_-)$. Here in this work, we try to provide a direct way to compute the matter field entropy, which definitely matches the earlier result. \\
\section{Brief discussion on the black hole in Jackiw-Teitelboim gravity coupled to a flat thermal bath}
\begin{figure}[!h]
    \centering
    \includegraphics[width=0.7\linewidth]{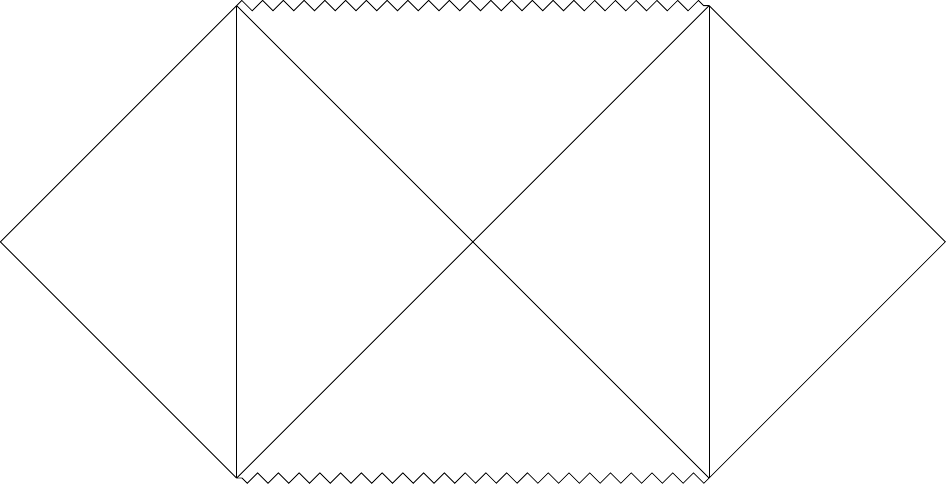}
    \caption{\noindent Penrose diagram of an eternal black hole in JT gravity with a flat auxiliary thermal bath system.}
    \label{fig1}
\end{figure}

\noindent In this section, we consider black hole solutions in two dimensions, that is, in $(1+1)$-dimensional gravitational theory \cite{Teitelboim:1983ux,Jackiw:1984je}. It is also to be noted that, in two dimensions, the Einstein-Hilbert part of the gravity action becomes topological. Therefore, to get a nontrivial gravitational dynamics, we need to couple a dilaton field with the gravity theory.
 %In this section we have briefly discussed black hole in JT gravity \textbf{REF} which is coupled to a pair of flat thermal baths. JT gravity is one of the solution of two dimensional dilatonic gravity. Here we have also assumed that the whole spacetime is filled with two dimensional conformal field.
The action for the two-dimensional dilatonic gravity theory is given by \cite{Mandal:1991tz,Grumiller:2002nm}
\begin{align}
    \mathcal{S}_{2d}&=\frac{1}{16\pi G_N}\int_{\mathcal{M}}d^2x\sqrt{-g}\Big[\phi R + U(\phi)(\nabla\phi)^2 + V(\phi)\Big]\nonumber\\
    &+\frac{1}{8\pi G_N}\int_{\partial \mathcal{M}}dx\sqrt{-\gamma}\phi \mathcal{K}~.
\end{align}
Here, we are not considering the dynamics of the dilaton. We get different gravity solutions from the above action for different choices of the $U(\phi)$ and $V(\phi)$. In particular, if we choose $U(\phi)=0$ and $V(\phi)=\frac{2\phi}{l^2}$, we get the black hole solution in  JT gravity. Here, $\phi$ represents the dilaton field profile, and it is given by $\phi(r)=\frac{r}{l}$. Before proceed further, it should be noted that the black hole is coupled with a flat thermal bath, and the full spacetime is filled with a two-dimensional conformal field theory. We should also keep in mind that there is coupling between the conformal matter field with the dilaton.
The JT gravity black hole solution of the above theory in the Schwarzschild gauge reads 
\begin{equation}
    ds^2 =-f(r)dt^2 +\frac{dr^2}{f(r)}~;~f(r)=\frac{(r^2 -r_h^2)}{l^2}~.
\end{equation}
In Kruskal coordinates, the metric inside the AdS spacetime reads
\begin{equation}\label{eq4}
    ds^2=-\mathcal{F}^2(r)dudv=-\frac{f(r)}{\kappa^2 uv}
\end{equation}
%\textbf{Figure}\\
The flat bath metric is given by the following form
\begin{equation}\label{eq5}
    ds^2_{Bath}=-\mathcal{F}^2(\xi,r)dudv=-\frac{f(\xi)}{\kappa^2 uv}
\end{equation}
The Kruskal coordinates for the right wedge (RW) read
\begin{align}
    &u=-e^{-\kappa(t-r^{*}(r))}\nonumber\\
    &v=e^{\kappa(t+r^{*}(r))}~.
\end{align}
Similarly, for the left wedge
\begin{align}
    &u=e^{\kappa(t+r^{*}(r))}\nonumber\\
    &v=-e^{-\kappa(t-r^{*}(r))}~.
\end{align}
where $\kappa=\frac{r_+}{l^2}$ is the surface gravity. In Fig.\eqref{fig1} we have represented the Penrose diagram of black hole in JT gravity attached with flat thermal bath.
The Hawking temperature $(T_H)$ and the thermal entropy of the black hole $(S_{BH})$ in this case is \textcolor{blue}{given by}
\begin{equation}
T_{H}=\frac{r_+}{2\pi l^2}~~;~~ S_{BH}=\frac{r_+}{4 G_{N}l}~.
\end{equation}
%\begin{widetext}
%~
      
%\end{widetext}
\section{Computation of mutual information between two different sets of sub-regions}
\begin{figure}
    \centering
    \includegraphics[width=0.7\linewidth]{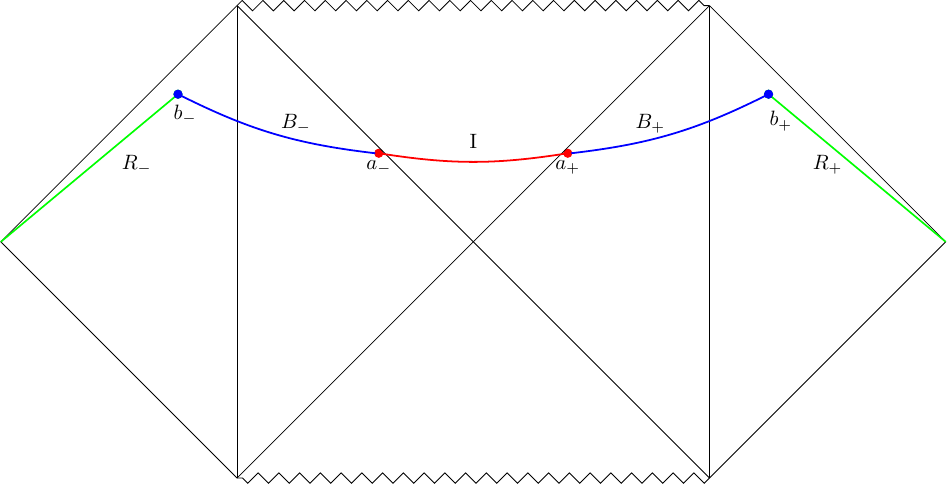}
    \caption{Penrose diagram of an eternal black hole in Jackiw-Teitelboim gravity with a flat auxiliary thermal bath system. 
    %The orange regions denote flat thermal baths. 
    The $R_{\pm}$ regions are denoted by green lines with the inner boundaries $b_{\pm}=(\pm t_b,b)$. The island region is shown in red with boundaries $a_{\pm}=(\pm t_a,a)$. The inner boundaries of $B_{\pm}$ regions are $a_{\pm}=(\pm t_a,a)$ and the outer boundaries are $b_{\pm}=(\pm t_b,b)$.
    }
    \label{fig2}
\end{figure}

\noindent In this section, we study the mutual information between two different sets of sub-regions lying on the same Cauchy slice. We should keep in mind that the entire analysis is done in the post Page time domain, that is, in the presence of the island region. We start our analysis by computing the mutual information between $B_+$ and $B_-$, that is, $I(B_{+}: B_{-})$. The regions $B_+$ and $B_{-}$ are shown by a blue line in the Penrose diagram given in Fig.\eqref{fig2}. The boundary region $B_{\pm}$ is denoted by $(b_{\pm}\rightarrow{a_{\pm}})$, where $a_{\pm}=(\pm t_a,a)$, represents the boundary of the island, and $b_{\pm}=(\pm t_b,b)$. 
%\begin{widetext}

%\end{widetext}
\noindent In our earlier works \cite{Saha:2021ohr,RoyChowdhury:2022awr,RoyChowdhury:2023eol,Saha:2025ttj,RoyChowdhury:2026jjp}, we have shown that the saturation of $I(B_{+}: B_{-})$ leads to the time-independent expression of the fine-grained entropy of Hawking radiation. This results in the correct form of Page curve for an eternal black hole. Here we shall look at the vanishing condition of $I(B_{+}: B_{-})$ once again. Let us first compute the mutual information between $B_+$ and $B_-$. We can compute $I(B_+: B_-)$ by the following formula
\begin{equation}\label{eq9}
    I(B_+:B_-)=S_{vN}(B_+)+S_{vN}(B_-)-S_{vN}(B_+\cup B_{-})
\end{equation}
where the expression of $S_{vN}(B_{\pm})$ and $S_{vN}(B_+\cup B_-)$ are given
\vspace{0.5cm}
\\
by \cite{Calabrese:2009qy}
\begin{align}\label{10}
   S_{vN}(B_{\pm})&=\frac{c}{3}\log(d(a_{\pm},b_{\pm}))\nonumber\\
   S_{vN}(B_{+}\cup B_{-})&=\frac{c}{3}\log\Bigg[\frac{d(a_{+},a_{-})d(b_{+},b_{-})d(a_{+},b_{+})d(a_{-},b_{-})}{d(a_{+},b_{-})d(a_{-},b_{+})}\Bigg]~.
\end{align}
Now using the above results in eq.\eqref{eq9}, we get the following result
\begin{equation}\label{Id}
 I(B_+:B_-)=\frac{c}{3}\log\left[\frac{d(a_+,b_-)d(a_-,b_+)}{d(a_+,a_-)d(b_+,d_-)}\right]~.
\end{equation}
Now one can compute the different distances that appear in the above formula by using the black hole and bath metrics given in eq.\eqref{eq4} and eq.\eqref{eq5} respectively.
\begin{widetext}
\noindent The expressions of the various distances that appear in the above results read
\begin{align}
d(a_+,b_+)&=\sqrt{2F(a)F(\xi,b)e^{\kappa(r^{*}(b)+r^{*}(a))}}\Big[\cosh[\kappa(r^{*}(a)-r^{*}(b))]-\cosh[\kappa(t_{a}-t_{b})]\Big]^{\frac{1}{2}}=d(a_{-},b_{-})\label{neweq6}\\
d(a_-,b_+)&=\sqrt{2F(a)F(\xi,b)e^{\kappa(r^{*}(b)+r^{*}(a))}}\Big[\cosh[\kappa(r^{*}(a)-r^{*}(b))]+\cosh[\kappa(t_{a}+t_{b})]\Big]^{\frac{1}{2}}=d(a_{+},b_{-})\label{neweq7}\\
d(b_{+},b_{-})&=2F(\xi,b)e^{\kappa r^{*}(b)}\cosh(\kappa t_{b})\label{neweq8}\\
d(a_{+},a_{-})&=2F(a)e^{\kappa r^{*}(a)}\cosh(\kappa t_{a})~.\label{neweq9}
\end{align}    
\end{widetext}
Now using the above results in eq.\eqref{Id}, we get
\begin{equation}\label{IBpm}
 I(B_+:B_-)=\log\left[\frac{d^2(a_+,b_-)}{d(a_+,a_-)d(b_+,b_-)}\right]~.
\end{equation}
The above expression suggests that the mutual information between $B_+$ and $B_-$ becomes zero if 
\begin{equation}
d^2(a_+,b_-)=d(a_+,a_-)d(b_+,b_-)~.
\end{equation}
Now using the results given in eq.\eqref{neweq7}, eq.\eqref{neweq8} and eq.\eqref{neweq9} in the above expression, we get \cite{Saha:2021ohr,RoyChowdhury:2022awr,RoyChowdhury:2023eol,Saha:2025ttj,RoyChowdhury:2026jjp}
\begin{equation}\label{scrambling time}
 t_a -t_b = \mid r^{*}(a) -r^{*}(b)\mid~.
\end{equation}
Before proceeding further, we would like to make a few comments regarding the above analysis. The above discussion shows that the mutual information between the regions $B_+$ and $B_-$ vanishes when $ t_a -t_b = \mid r^{*}(a) -r^{*}(b)\mid\equiv t_{Scr}$. This implies that the correlation between the regions $B_+$ and $B_-$ vanishes at the scrambling time. This in turn means that the entanglement wedge associated with $B_+$ and $B_-$ becomes disconnected at the scrambling time.\\
Now we proceed to compute the mutual information between the other two sets of regions on the Cauchy slice. In particular, we compute the mutual information between the island region $\mathcal{I}$ and the region outside the black hole where the radiation is collected, that is, the region $R=R_+\cup R_-$. The mutual information between $\mathcal{I}$ and $R~(=R_+\cup R_-)$ can be computed by using the following expression
\begin{equation}\label{IR}
I(\mathcal{I}:R)=S_{vN}(\mathcal{I})+S_{vN}(R)-S_{vN}(\mathcal{I} \cup R)~.
\end{equation}
Now, from the pure state property of the von-Neumann entropy, we can write $S_{vN}(\mathcal{I}\cup R)=S_{vN}(\mathcal{I}\cup R_{+}\cup R_{-})=S_{vN}(B_{+}\cup B_{-})$. Therefore, the above expression takes the following form
\begin{equation}\label{mutual information formula}
I(\mathcal{I}:R)=S_{vN}(\mathcal{I})+S_{vN}(R)-S_{vN}(B_{+}\cup B_{-})~.
\end{equation}
The von-Neumann entropy of the matter fields in the region $R$ is given by 
\begin{equation}
    S_{vN}(R)=S_{vN}(R_{+}\cup R_{-})=S_{vN}(R^{c})=\frac{c}{3}\log d(b_{+},b_{-})~.
\end{equation}
Similarly, for the island region ($\mathcal{I}$), the von-Neumann entropy reads
\begin{equation}
    S_{vN}(\mathcal{I})=\frac{c}{3}\log d(a_{+},a_{-})~.
\end{equation}
\\
As we are using free 2D CFTs in the matter sector, the von-Neumann entropy for the region $B_{+}\cup B_{-}$ can be computed using the formula for disjoint regions \cite{Calabrese:2009ez} given in eq.\eqref{10}. 
%\begin{equation}
  %  S_{vN}(B_{+}\cup B_{-})=\frac{c}{3}\log\Bigg[\frac{d(a_{+},a_{-})d(b_{+},b_{-})d(a_{+},b_{+})d(a_{-},b_{-})}{d(a_{+},b_{-})d(a_{-},b_{+})}\Bigg]~.
%\end{equation}
Now substituting the results of $S_{vN}(R)$,~$S_{vN}(\mathcal{I})$ and $S_{vN}(\mathcal{I}\cup R)=S_{vN}(B_+\cup B_-)$ in eq.\eqref{IR}, we get the following expression of mutual information between the regions $\mathcal{I}$ and $R$
\begin{equation}\label{MId}
I(\mathcal{I}:R)=\frac{2c}{3}\log\Bigg[\frac{d(a_{+},b_{-})}{d(a_{+},b_{+})}\Bigg]~.
\end{equation}
Now substituting the expressions of $d(a_{+},b_{+})$ (given in eq.\eqref{neweq6}) and $d(a_{+},b_{-})$(given in eq.\eqref{neweq7}) in the above result, we get

%Following \cite{RoyChowdhury:2022awr}, we already have the expressions for the distances $d(a_{+},b_{-})$ and $d(a_{+},b_{+})$, this gives
%\begin{widetext}
\begin{equation}\label{MI I:R}
    I(\mathcal{I}:R)=\frac{c}{3}\log\Bigg[\frac{\cosh [\kappa(r^{*}(a)-r^{*}(b))]+\cosh [\kappa (t_a +t_b)]}{\cosh [\kappa(r^{*}(a)-r^{*}(b))]-\cosh [\kappa (t_a -t_b)]}\Bigg]~.
\end{equation}
The above expression suggests that the mutual information between $\mathcal{I}$ and $R~(=R_+\cup R_-)$ can never be zero. This implies that the entanglement wedge of $\mathcal{I}\cup R$ is always in the connected phase. This, in turn, implies that these regions are always correlated in contrast to our earlier discussion.\\
% \begin{figure}
%     \centering
%     \includegraphics[width=\linewidth]{MI.pdf}
%     \caption{Variation of the mutual information between the regions $\mathcal{I}$ and $R$ (that is $I(\mathcal{I}:R)$) with respect to a parameter $\epsilon$. The plot is done for $r_+=1$, $a=2r_+$, $b=10r_+$ and $t_a=2$.}
%     \label{fig:MI plot}
% \end{figure}
\begin{widetext}

\begin{figure}
    \centering

    \begin{subfigure}{0.48\linewidth}
        \centering
        \includegraphics[width=\linewidth]{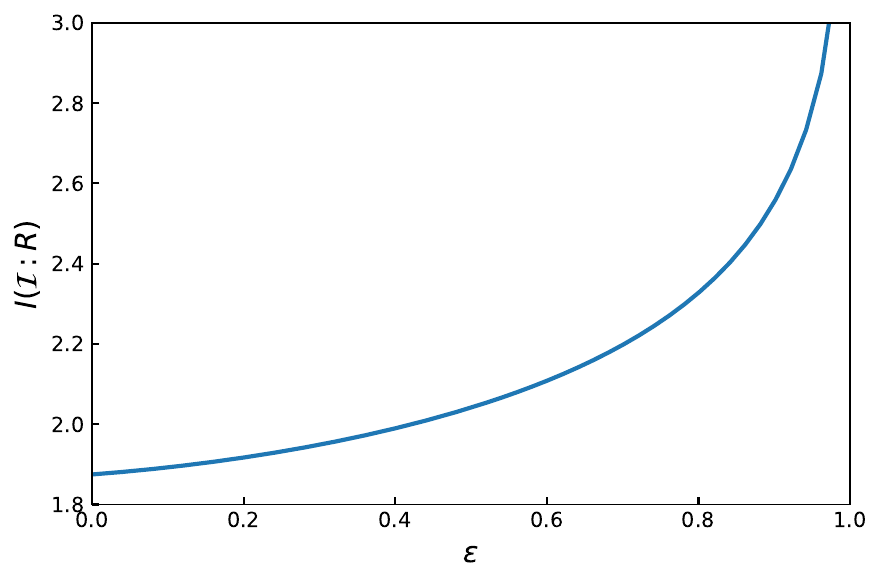}
        \caption{$I(\mathcal{I}: R)$ vs $\epsilon$}
        \label{fig:MI plot}
    \end{subfigure}
    \hfill
    \begin{subfigure}{0.48\linewidth}
        \centering
        \includegraphics[width=\linewidth]{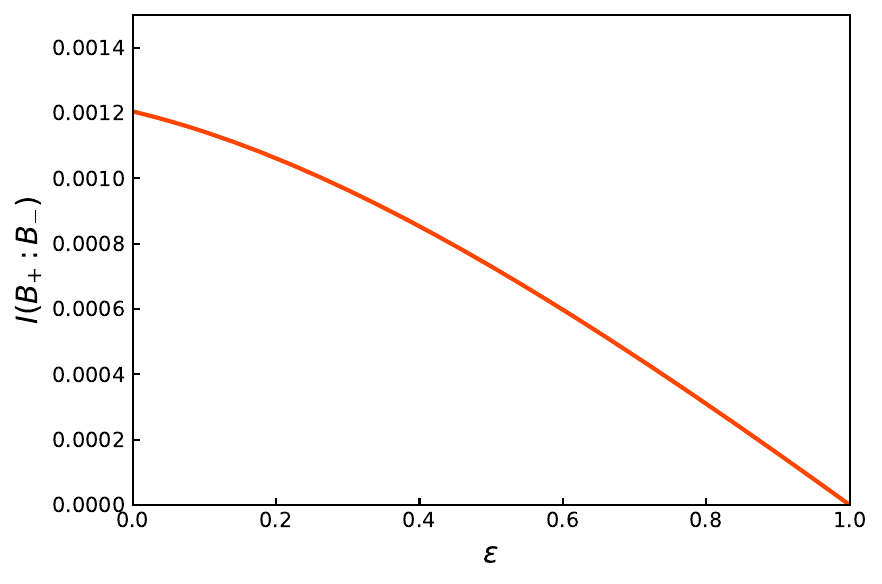}
        \caption{$I(B_+: B_-)$ vs $\epsilon$}
        \label{fig:MIB}
    \end{subfigure}

    \caption{The left panel shows the variation of the mutual information between $\mathcal{I}$, $R$ ($I(\mathcal{I}: R)$) with respect to $\epsilon$. The right panel shows the variation of $I(B_+: B_-)$ with respect to $\epsilon$. Both the plots are done for $r_+=1$, $a=2r_+$, $b=10r_+$ and $t_a=2$.}
    \label{fig:comparison MI}
\end{figure}
\end{widetext}
%\end{widetext}
%In order to obtain the above expression, we have used the following forms for $d(a_{+},b_{-})$ and $d(a_{+},b_{+})$
%\begin{widetext}
    %\begin{align}
        %d(a_{+},b_{-})&=\sqrt{2 \mathcal{F}(a)\mathcal{F}(\xi,b)e^{\kappa ( r^{*}(a)+r^{*}(b))}\Big(\cosh [\kappa(r^{*}(a)-r^{*}(b))]+\cosh [\kappa (t_a +t_b)]\Big)}\nonumber\\
       % d(a_{+},b_{+})&=\sqrt{2 \mathcal{F}(a)\mathcal{F}(\xi,b)e^{\kappa ( r^{*}(a)+r^{*}(b))}\Big(\cosh [\kappa(r^{*}(a)-r^{*}(b))]-\cosh [\kappa (t_a -t_b)]\Big)}
   % \end{align}
%\end{widetext}
%After substituting the above expressions for $d(a_{+},b_{-})$ and $d(a_{+},b_{+})$ in eq.\eqref{MI in d}, we get the following result 
As discussed in the previous studies, at the scrambling time, the mutual information between the two regions $B_{+}$ and $B_{-}$ vanishes. Therefore, one can write the following
\begin{equation}
    I(B_+: B_-)\mid_{ t_a -t_b = \mid r^{*}(a) -r^{*}(b)\mid~}=0~. 
\end{equation}
%Several earlier studies \cite{Saha:2021ohr,RoyChowdhury:2022awr,RoyChowdhury:2023eol}  have also revealed that the Scrambling time \cite{Sekino:2008he,Hayden:2007cs} of the black hole in terms of the island parameter, which reads
%\begin{equation}
    %t_{scr}=t_a -t_b = \mid r^{*}(a) -r^{*}(b)\mid~.
%\end{equation}
We now make a crucial observation. If we impose the condition of vanishing mutual information between $B_+$ and $B_-$ on $I(\mathcal{I}\cup R)$, the mutual information between $\mathcal{I}$ and $R$ diverges.
Now from eq.\eqref{MI I:R}, we can clearly observe that at the scrambling time, the mutual information between the regions $\mathcal{I}$ and $R$ diverges. Thus, we have the following
\begin{equation}
    I(\mathcal{I}:R)\mid_{ t_a -t_b = \mid r^{*}(a) -r^{*}(b)\mid}\to \infty~.
\end{equation}
Some important observations in this regard are in order. It is very interesting to note that when the mutual information between the regions $B_{+}$ and $B_{-}$ vanishes, the mutual information between the island region ($I$) and the region outside the black hole event horizon ($R$) blows up. This phenomenon clearly indicates the distribution of quantum entanglement in different regions of the Cauchy slice. We would like to mention another important point regarding the mutual information between $\mathcal{I}$ and $R$. Since mutual information between disjoint regions is UV finite by construction, the divergence in the mutual information does not originate from an ultraviolet divergence of the individual entanglement entropies appearing in the expression of $I(\mathcal{I}:R)$. The nontrivial feature of the present setup is that this singular behaviour does not arise by bringing the two endpoints of the regions (island and the radiation region) together; rather, it follows from the condition of vanishing of mutual information between $B_+$ and $B_-$, namely,  $I(B_{+}:B_{-})=0$, which leads to $t_{a}-t_{b}=|r_{*}(a)-r_{*}(b)|$. This is evident from the plot given in Fig.\eqref{fig:MI plot}. Hence, the divergence is physically meaningful.\\
We have also graphically represented our result of mutual information (eq.\eqref{MI I:R}) between the island region ($\mathcal{I}$) and the region outside the black hole horizon ($R$) for a fixed value of island time ($t_a$) in Fig.\eqref{fig:MI plot}. In order to do the plot, we have introduced a parameter $\epsilon~ (0<\epsilon\le1)$,  such that the difference between the island time ($t_a$) and the observer's time ($t_b$) can be, in general, written as
\begin{equation}
    t_a-t_b=\epsilon\Big[r^{*}(a) -r^{*}(b)\Big]~.
\end{equation}
The above equation reduces to the usual expression for the scrambling time (in eq.\eqref{scrambling time}) when $\epsilon=1$. In Fig.\eqref{fig:MI plot}, we can clearly see that at the scrambling time (that is, at $\epsilon=1$), the regions $\mathcal{I}$ and $R$ become maximally entangled and the mutual information between them diverges. We have also expressed the analytical expression for the mutual information between the regions $B_+$ and $B_-$ (in eq.\eqref{IBpm}) with respect to the parameter $\epsilon$ in Fig.\eqref{fig:MIB}. Fig.\eqref{fig:MIB} clearly shows that at $\epsilon=1$, $I(B_+:B_-)$ becomes zero, and this is precisely the point where $I(\mathcal{I}:R)$ diverges. It is also evident from the plots that as $\epsilon$ increases the mutual information between $B_+$ and $B_-$, that is, $I(B_+:B_-)$ decreases. On the other hand the mutual information between $\mathcal{I}$ and $R ~(R_+\cup R_-)$, $I(\mathcal{I}:R)$, increases with increase in $\epsilon$. Finally at $\epsilon=1$, which corresponds to the condition $t_{a}-t_{b}=|r_{*}(a)-r_{*}(b)|$, the mutual information between the exterior regions ($B_{+}$ and $B_{-}$) vanishes while the mutual information between the island $(\mathcal{I})$ and radiation ($R$) diverges. 
This clearly shows the phenomenon of redistribution of geometric entanglement between different regions on the Cauchy slice. In making these plots, we have chosen $r_+=1$, $a=2r_+$, $b=10r_+$ and $t_a=2$.
\\
Before proceeding further, we would like to make a few comments regarding the above finding. In our earlier works, we have shown that just after the Page time, the island leads to the saturation of mutual information between $B_+$ and $B_-$. This, in turn, results in a time-independent expression of von-Neumann entropy of the matter field appearing in the island formula. Thus, we get the correct form of the Page curve in the post-Page time domain. In contrast, our present work suggests that the mutual information between $\mathcal{I}$ and $R$ diverges under the very conditions for which $I(B_+ : B_-)$ vanishes. Therefore, we can reformulate our previous idea in terms of the mutual information between the island region $(\mathcal{I})$ and bath region $(R)$. Our results suggest that just after the Page time, the inclusion of the island gives rise to a divergent mutual information between $\mathcal{I}$ and $R$. It can be shown that substituting the condition for the divergence of $I(\mathcal{I}:R)$ into the relation
$S_{vN}(\mathcal{I}\cup R)=S_{vN}(B_{+}\cup B_{-})$,
yields the desired time-independent expression for the von-Neumann entropy of the matter field. Furthermore, by applying the standard extremization procedure of the island formula, we obtain the location of the island endpoint and the corresponding time-independent expression for the von-Neumann entropy of the Hawking radiation.\\
Therefore, our analysis strongly suggests that the distribution of quantum entanglement in a Cauchy slice plays a crucial role in getting the Page curve in the post-Page time domain. It is worth emphasizing that both $I(B_{+}:B_{-})=0$ and $I(\mathcal{I}:R)\to\infty$ imply the same condition, indicating a nontrivial correspondence between the vanishing of the mutual information between $B_{+}$ and $B_{-}$ and the divergence of the mutual information between $\mathcal{I}$ and $R$. Hence, both of them lead to the correct Page curve. Therefore, the island formula can be written in the following form
\begin{align}
    S(R)&=\min \underset{\mathcal{I}}{\text{ext}} \left\{\frac{A(\partial \mathcal{I})}{4G_N}+S_{vN}(\mathcal{I}\cup R)\right\}\nonumber\\
    &=\min \underset{\mathcal{I}}{\text{ext}} \left\{\frac{A(\partial \mathcal{I})}{4G_N}+S_{vN}(B_+\cup B_-)|_{I(B_+ : B_-)=0}\right\}\nonumber\\
    &=\min \underset{\mathcal{I}}{\text{ext}} \left\{\frac{A(\partial \mathcal{I})}{4G_N}+S_{vN}(B_{+}\cup B_-)|_{I(\mathcal{I}:R)\to\infty}\right\}~.
\end{align}
The above result has an important implication for the emergence of the Page curve. In particular, reproducing the expected Page curve requires satisfying at least one of the two conditions. Either the subsystems $B_{+}$ and $B_{-}$ must be completely uncorrelated, such that their mutual information vanishes,
\begin{equation}
I(B_{+}:B_{-})=0,
\end{equation}
or the island $\mathcal{I}$ must share an infinite amount of mutual information with the Hawking radiation region $R=R_{+}\cup R_{-}$, namely,
\begin{equation}
I(\mathcal{I}:R)=\infty.
\end{equation}
The former condition implies the absence of correlations between the two subsystems $B_{+}$ and $B_{-}$, while the latter indicates an infinitely strong entanglement between the island and the radiation degrees of freedom. Therefore, the emergence of the correct Page curve is intimately tied either to the complete decoupling of $B_{+}$ and $B_{-}$ or to an infinitely strong correlation between the island and the radiation sector. The physical interpretation of this divergence in $I(\mathcal{I}:R)$ can be thought of as the complete correlation redistribution at the scrambling time.
\\ \\
\section{Computation of Tripartite information between :~\texorpdfstring{$\mathrm{I(\mathcal{I}:R_+:R_-)}$}{I(I:R+:R-)}}
\begin{widetext}

\begin{figure}
    \centering

    \begin{subfigure}{0.48\linewidth}
        \centering
        \includegraphics[width=\linewidth]{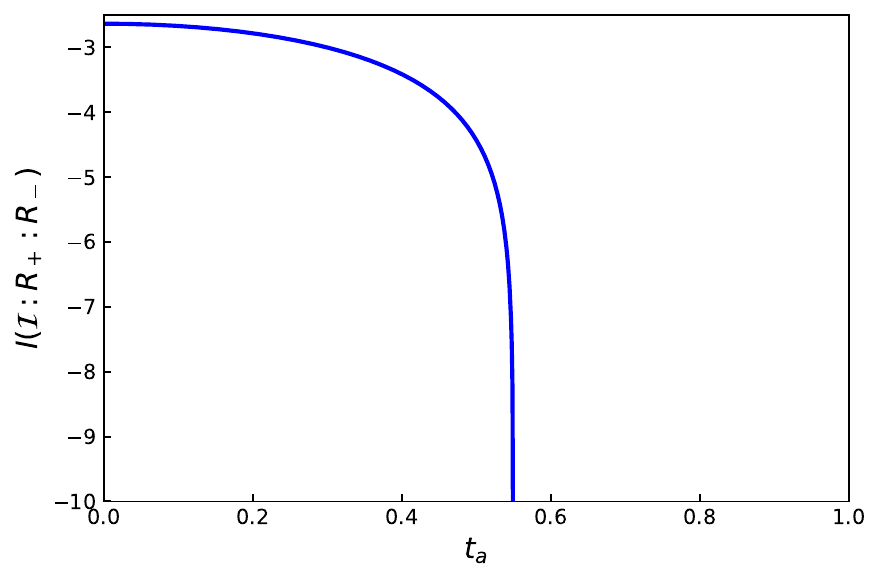}
        \caption{$I(\mathcal{I}: R_+: R_-)$ vs $t_a$}
        \label{fig:Ita}
    \end{subfigure}
    \hfill
    \begin{subfigure}{0.48\linewidth}
        \centering
        \includegraphics[width=\linewidth]{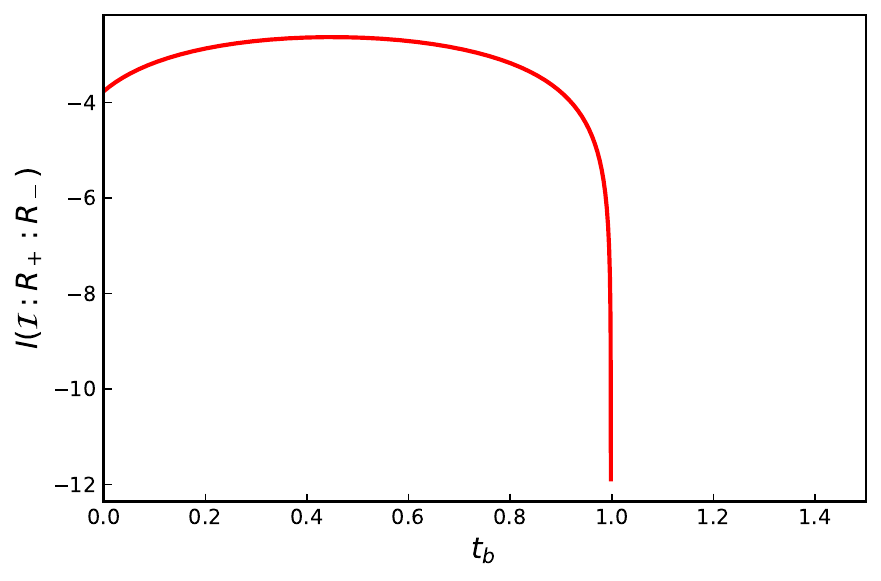}
        \caption{$I(\mathcal{I}: R_+: R_-)$ vs $t_b$}
        \label{fig:Itb}
    \end{subfigure}

    \caption{The left panel shows the variation of the tripartite information between $\mathcal{I}$, $R_+$ and $R_-$ ($I(\mathcal{I}: R_+: R_-)$) with respect to the island time $t_a$. The right panel shows the variation of $I(\mathcal{I}: R_+: R_-)$ with respect to the observer's time $t_b$. Both of the plots are done for $r_+=1$, $a=2r_+$ and $b=10r_+$.}
    \label{fig:comparison}
\end{figure}
\end{widetext}

\noindent In this section, we will compute the tripartite information between three disjoint regions on the Cauchy slice ($\mathcal{I}, R_+$ and $R_-$). We will start with the von-Neumann entropy for the three disjoint regions $\mathcal{I}$, $R_+$ and $R_-$, which is given by the following relation \cite{Chuang:2000}
\begin{widetext}
\begin{equation}
    S_{vN}(\mathcal{I}\cup R_+\cup R_-)=S_{vN}(\mathcal{I})+S_{vN}(R_+)+S_{vN}(R_-)-S_{vN}(\mathcal{I}\cap R_+)-S_{vN}(\mathcal{I}\cap R_-)-S_{vN}(R_+\cap R_-)+S_{vN}(\mathcal{I}\cap R_+\cap R_-)~.
\end{equation}
\end{widetext}
We already know that for two subsystems $S_{vN}(A\cap B)$ is nothing but the mutual information between subsystems $A$ and $B$. We would also like to mention that the symbol $\cap$ denotes the informational analogy of intersection between two regions, not the geometric intersection. Hence, the above equation can be recast in the following form
\begin{widetext}
\begin{equation}
    S_{vN}(\mathcal{I}\cup R_+\cup R_-)=S_{vN}(\mathcal{I})+S_{vN}(R_+)+S_{vN}(R_-)-I(\mathcal{I}: R_+)-I(\mathcal{I}: R_-)-I(R_+: R_-)+I(\mathcal{I}: R_+: R_-)
\end{equation}
\end{widetext}
where $I(\mathcal{I}: R_+: R_-)$ is the tripartite information between the island region $\mathcal{I}$ and the radiation regions $R_+$ and $R_-$ respectively. We can further simplify the above equation by rewriting the expressions of mutual information in terms of von-Neumann entropies, which reads
\begin{widetext}
\begin{equation}\label{tri part 1}
    S_{vN}(\mathcal{I}\cup R_+\cup R_-)=S_{vN}(\mathcal{I}\cup R_+)+S_{vN}(\mathcal{I}\cup R_-)+S_{vN}(R_+\cup R_-)-S_{vN}(\mathcal{I})-S_{vN}(R_+)-S_{vN}(R_-)+I(\mathcal{I}: R_+: R_-)~.
\end{equation}  
\end{widetext}
Now, we will use the respective expressions for all the terms in the above equation, along with the property of the Cauchy slice in order to obtain an analytical form for $I(\mathcal{I}: R_+: R_-)$. For individual regions $\mathcal{I}$, $R_+$ and $R_-$ the von-Neumann entropies are given by the following expressions respectively
\begin{align}
    &S_{vN}(\mathcal{I})=\frac{c}{3}\log d(a_+,a_-)\\
    &S_{vN}(R_+)=\frac{c}{3}\log d(b_+,e_+)\\
    &S_{vN}(R_-)=\frac{c}{3}\log d(b_-,e_-)~.
\end{align}
For two disjoint regions $\mathcal{I}$ and $R_+$ the von-Neumann entropy is given by the following formula \cite{Calabrese:2009ez}
\begin{equation}
    S_{vN}(\mathcal{I}\cup R_+)=\frac{c}{3}\log\Bigg[\frac{d(a_+,b_+)d(a_-,e_+)d(b_+,e_+)d(a_+,a_-)}{d(a_+,e_+)d(a_-,b_+)}\Bigg]~.
\end{equation}
Using the symmetry property on the Cauchy slice, one can easily obtain the expression of the von-Neumann entropy between $\mathcal{I}$ and $R_-$. From symmetry property, we can tell that $S_{vN}(\mathcal{I}\cup R_+)=S_{vN}(\mathcal{I}\cup R_-)$. \\
With all of these expressions in hand, we will proceed to rewrite eq.\eqref{tri part 1}, which gives
\begin{widetext}
    \begin{align}
    S_{vN}(\mathcal{I}\cup R_+\cup R_-)&=\frac{2c}{3}\log\Bigg[\frac{d(a_+,b_+)d(a_-,e_+)d(b_+,e_+)d(a_+,a_-)}{d(a_+,e_+)d(a_-,b_+)}\Bigg]-\frac{c}{3}\log d(a_+,a_-)-\frac{c}{3}\log d(b_+,e_+)\nonumber\\
    &-\frac{c}{3}\log d(b_-,e_-)+\frac{c}{3}\log d(b_+,b_-)+ I(\mathcal{I}: R_+: R_-)~.
\end{align}
\end{widetext}
After carefully adjusting the above terms and using the symmetry of different distances on the Cauchy slice, we obtain
\begin{widetext}
\begin{align}
    S_{vN}(\mathcal{I}\cup R_+\cup R_-)&=\frac{c}{3}\log d(a_+,a_-)+\frac{c}{3}\log d(b_+,b_-)+\frac{c}{3}\log [d(a_+,b_+)d(a_-,b_-)]+\frac{c}{3}\log [d(a_-,e_+)d(a_+,e_-)]\nonumber\\&-\frac{c}{3}\log [d(a_+,e_+)d(a_-,e_-)]-\frac{c}{3}\log [d(a_-,b_+)d(a_+,b_-)]+ I(\mathcal{I}: R_+: R_-)~.
\end{align}  
\end{widetext}
With a little bit of algebra, the above expression for $S_{vN}(\mathcal{I}\cup R_+\cup R_-)$ can be further simplified and can be recast in the following form
\begin{widetext}
    \begin{align}\label{S last}
    S_{vN}(\mathcal{I}\cup R_+\cup R_-)&=\frac{c}{3}\log\Bigg[\frac{d(a_+,a_-)d(b_+,b_-)d(a_+,b_+)d(a_-,b_-)d(a_-,e_+)d(a_+,e_-)}{d(a_+,e_+)d(a_-,e_-)d(a_-,b_+)d(a_+,b_-)}\Bigg]+I(\mathcal{I}: R_+: R_-)\nonumber\\
    &=\frac{c}{3}\log\Bigg[\frac{d(a_{+},a_{-})d(b_{+},b_{-})d(a_{+},b_{+})d(a_{-},b_{-})}{d(a_{+},b_{-})d(a_{-},b_{+})}\Bigg]+\frac{c}{3}\log\Bigg[\frac{d(a_+,e_-)d(a_-,e_+)}{d(a_+,e_+)d(a_-,e_-)}\Bigg]+I(\mathcal{I}: R_+: R_-)~.
\end{align}
\end{widetext}
From eq.\eqref{S last}, we can easily identify that the first term in the above equation is nothing but $S_{vN}(B_+\cup B_-)$. Again, from the property of the Cauchy slice, we have $S_{vN}(\mathcal{I}\cup R_+\cup R_-)=S_{vN}(B_+\cup B_-)$. Therefore, we can write 
\begin{align}
    &\frac{c}{3}\log\Bigg[\frac{d(a_+,e_-)d(a_-,e_+)}{d(a_+,e_+)d(a_-,e_-)}\Bigg]+I(\mathcal{I}: R_+: R_-)=0\nonumber\\
\end{align}
Solving the above equation for $I(\mathcal{I}: R_+: R_-)$, we finally get
\begin{align}\label{tri part 2}
    I(\mathcal{I}: R_+: R_-)&=\frac{c}{3}\log\Bigg[\frac{d(a_+,e_+)d(a_-,e_-)}{d(a_+,e_-)d(a_-,e_+)}\Bigg]~.
   % &=\frac{2c}{3}\log\Bigg[\frac{d(a_+,e_+)}{d(a_+,e_-)}\Bigg]~.
\end{align}
In order to get an explicit form $ I(\mathcal{I}: R_+: R_-)$, we have to substitute the expressions of various distances appearing in the above result. We can compute these distances by using the metric given in eq.\eqref{eq4} and eq.\eqref{eq5}. Therefore, we get the following expressions of different distances appearing in the above result of $I(\mathcal{I}:R_+:R_-)$
\begin{widetext}
\begin{align}
d(a_{+},e_{+})&=\sqrt{2 \mathcal{F}(a)\mathcal{F}(\xi,e)e^{\kappa ( r^{*}(a)+r^{*}(e))}\Big(\cosh [\kappa(r^{*}(e)-r^{*}(a))]-\cosh \kappa t_a \Big)}=d(a_{-},e_{-})\nonumber\\
d(a_{+},e_{-})&=\sqrt{2 \mathcal{F}(a)\mathcal{F}(\xi,e)e^{\kappa ( r^{*}(a)+r^{*}(e))}\Big(\cosh [\kappa(r^{*}(e)-r^{*}(a))]+\cosh \kappa t_a \Big)}=d(a_{-},e_{+})~.
\end{align}
\end{widetext}
The above result suggests that $d(a_+,e_+)=d(a_-,e_-)$ and $d(a_+,e_-)=d(a_-,e_+)$. This results in the following
\\ \\
\begin{equation}
   I(\mathcal{I}: R_+: R_-)= \frac{2c}{3}\log\Bigg[\frac{d(a_+,e_+)}{d(a_+,e_-)}\Bigg]~.
\end{equation}
Now we can substitute the expressions of $d(a_{+},e_{+})$ and $d(a_+,e_-)$ in the above result. This yields
\\ \\
\begin{equation}\label{tri final}
    I(\mathcal{I}: R_+: R_-)=\frac{c}{3}\log\Bigg[\frac{\cosh [\kappa(r^{*}(e)-r^{*}(a))]-\cosh \kappa t_a }{\cosh [\kappa(r^{*}(e)-r^{*}(a))]+\cosh \kappa t_a}\Bigg]~.
\end{equation}
We will now move forward to graphically represent our result of the tripartite information $I(\mathcal{I}: R_+: R_-)$ in eq.\eqref{tri final}. In Fig.\eqref{fig:Ita}, we have plotted $I(\mathcal{I}: R_+: R_-)$ with respect to island time $t_a$, and in Fig.\eqref{fig:Itb} we have shown the variation of the tripartite information with respect to the observer's time $t_b$. Both the plots are done for $r_+=1$, $a=2r_+$ and $b=10r_+$. The plots clearly show that the tripartite information between $\mathcal{I}$, $R_+$ and $R_-$ is always negative. Hence, it indicates the superextensive nature of the mutual information. Also, the tripartite information in this case is always monogamous in nature. It should be mentioned that in the context of quantum information theory, this monogamy property is related to the security of quantum cryptography, as, unlike classical correlation, quantum entanglement is not a shareable quantity.
The negativity of the tripartite mutual information
$I\left(\mathcal{I}:R_{+}:R_{-}\right)$
implies that the information contained in the island $\mathcal{I}$ is encoded nonlocally in the two radiation sectors. In particular,
\vspace{0.1cm}
\[
I\!\left(\mathcal{I}:R_{+}:R_{-}\right)
>
I\!\left(\mathcal{I}:R_{+}\right)
+
I\!\left(\mathcal{I}:R_{-}\right),
\]
% \vspace{0.1cm}
indicating that a portion of the information about $\mathcal{I}$ is accessible only through the joint correlations between $R_{+}$ and $R_{-}$. Thus, the island information is not redundantly stored in either radiation sector individually but is instead distributed in a genuinely multipartite manner across the full radiation system. This suggests that the reconstruction of the island degrees of freedom may require access to the combined radiation system $R_{+}R_{-}$, reflecting a highly nonlocal encoding of information.
%In order to obtain the above form of $I(\mathcal{I}: R_+: R_-)$, we have used the following expressions of $d(a_{+},e_{+})$ and $d(a_{+},e_{-})$ 

% \begin{figure}
%     \centering
%     \includegraphics[width=0.5\linewidth]{Ita.pdf}
%     \caption{Caption}
%     \label{fig:placeholder}
% \end{figure}
% \begin{figure}
%     \centering
%     \includegraphics[width=0.5\linewidth]{Itb.pdf}
%     \caption{Caption}
%     \label{fig:placeholder1}
% \end{figure}
%\end{widetext}
\section{Conclusion}
\noindent Now we will summarize our findings. In this present work, we have studied the mutual information between the island ($\mathcal{I}$) and the bath region $(R=R_+\cup R_-)$, that is, $I(\mathcal{I}:R)$. We have found that this mutual information can never vanish because it leads to a mathematical inconsistency. On the other hand, we have shown that $I(\mathcal{I}:R)$ diverges if $t_a-t_b=|r_{*}(a)-r_{*}(b)|$. This condition of diverging mutual information is quite surprising. This is exactly the same condition under which the mutual information between $B_+$ and $B_-$ vanishes. Therefore, our analysis suggests that under similar conditions, two different mutual information have different characteristics. In particular $I(B_+: B_-)$ vanishes and $I(R_+: R_-)$ diverges. This leads us to conjecture the concept of the distribution of geometric entanglement. In our earlier work we have shown that, just after the Page time, the vanishing of mutual information between $B_+$ and $B_-$ ($I(B_+: B_-)=0$) leads us to a time independent expression of von-Neumann entropy of the matter field ($S_{vN}(\mathcal{I}\cup R)$), in the post Page time domain. This results in the correct form of Page curve for an eternal black hole. On the other hand the mutual information between  ($\mathcal{I}$) and $(R=R_+\cup R_-)$, that is, $I(\mathcal{I}: R)$ diverges under the similar condition. Therefore, we can argue that, after the Page time, the divergence of $I(\mathcal{I}: R)$ leads to the correct time independent expression of $S_{vN}(\mathcal{I}\cup R)$ in the post Page time domain. Hence, obtaining the expected Page curve necessitates one of the following conditions: either the subsystems $B_{+}$ and $B_{-}$ are completely uncorrelated, $I(B_{+}:B_{-})=0$, or the island $\mathcal{I}$ is maximally correlated with the radiation region $R=R_{+}\cup R_{-}$, namely $I(\mathcal{I}:R)\rightarrow \infty$. This implies one can get the correct form of the Page curve by demanding either $I(B_{+}:B_{-})=0$ or $I(\mathcal{I}:R)\rightarrow \infty$. Therefore, for getting the correct form of the Page curve, either the regions $B_+$ and $B_-$ get completely disconnected, or the regions $\mathcal{I}$ and $R$ get strongly correlated. Therefore, there is an alternative approach to get the correct form of the Page curve by demanding  $I(\mathcal{I}:R)\rightarrow \infty$. This result may play an important role in obtaining the correct Page curve for an evaporating black hole. We leave a detailed investigation of this issue for future work. It would be interesting to determine whether these conditions are essential for reproducing the expected Page-curve behavior in a dynamical setting.\\
We have also computed the tripartite mutual information among the $\mathcal{I}$,$R_+$ and $R_-$, that is, $I(\mathcal{I}: R_+: R_-)$. Although the direct computation of $I(\mathcal{I}: R_+: R_-)$ is quite challenging, we have discussed a systematic way to compute $I(\mathcal{I}: R_+: R_-)$ using the properties of the Cauchy slice. We have also graphically represented the variation of $I(\mathcal{I}: R_+: R_-)$ with respect to the island time $t_a$ and the observer's time $t_b$. Both of the plots suggest that this tripartite mutual information is negative for all values of horizon radius. This implies that the tripartite mutual information, $I(\mathcal{I}: R_+: R_-)$, is monogamous and the mutual information has a superextensive nature. This monogamous nature suggests that quantum entanglement in this setting is not a shareable resource. In other words, we can say that the entangled correlations between $\mathcal{I}$ and one of the radiation regions (say $R_+$) cannot be shared with the other radiation region ($R_-$). Hence, the quantum correlation between $\mathcal{I}$ and $R_+$ is encrypted and cannot be shared in an exact form with the other radiation region $R_-$.
\section*{Acknowledgment}
\noindent ARC would like to thank SNBNCBS for the Bridge Fellowship. SP would like to thank SNBNCBS for the Senior Research Fellowship. The authors thank the anonymous referee for useful comments and suggestions.
\bibliographystyle{hephys.bst}
\bibliography{ref}
\end{document}